

\input jnl

\rightline{hepth/9304148}
\rightline{NSF-ITP-93-36}
\def\cI{{\cal I}}

\title NUMERICAL ANALYSIS OF BLACK HOLE EVAPORATION

\author Tsvi Piran\footnote{$^*$}{On leave from  Racah Institute of Physics,
The Hebrew University, Jerusalem, Israel}

\affil Center for Astrophysics, Harvard University
Cambridge, Ma 02138

\centerline{and}

\author Andrew Strominger

\affil Institute for Theoretical Physics
 and Department of Physics
University of California
 Santa Barbara, CA 93106-4030

\abstract{
 Black hole formation/evaporation in two-dimensional dilaton gravity can be
described, in the limit where the number $N$ of matter fields becomes large,
by a set of second-order
partial differential equations. In this paper we solve these
equations numerically. It is shown that, contrary to some previous suggestions,
black holes evaporate completely a finite time after formation. A boundary
condition is required to evolve the system beyond the naked singularity at the
evaporation endpoint. It is argued that this may be naturally chosen  so as to
restore the system to the vacuum. The analysis also applies to the low-energy
scattering of $S$-wave fermions by four-dimensional extremal, magnetic,
dilatonic black
holes.}
\endtitlepage
In recent work [\cite{CGHS}], referred to as CGHS, a set of second-order
partial differential equations were found to describe both $1+1$ dimensional
black hole formation/evaporation (including backreaction)
and scattering of $S$-wave matter by $3+1$ dimensional
extremal black holes. The CGHS equations apply to theories of gravity
 augmented by a scalar dilaton and $N$ matter
fields, and are the first term in a $1/N$ expansion of the full quantum theory.
They provide a concrete arena in which to address the puzzles of quantum
mechanical black holes. The CGHS equations apparently are not
analytically soluble. Attempts have been made to remedy this problem by
modifying the gravity-dilaton couplings. In [\cite{ABC}] modifications
were found which led to exactly soluble (even for finite $N$) quantum
theories. Unfortunately these theories are unphysical because the Hamiltonian
is unbounded from below [\cite{gs}]. This difficulty was circumvented in
[\cite{RST}], henceforth referred to as RST, by imposing boundary conditions
at the core of the black hole. These boundary conditions ruin the exact
solubility of the full quantum theory, but {\it not\/} of the large-$N$
semiclassical equations.

A fairly complete picture of black hole evaporation has emerged in the RST
model.
Black holes evaporate a finite time after formation, and the system can then
 be restored to the vacuum via  an endpoint boundary condition [\cite{RST,AS}].
Information is destroyed at the spacelike black hole singularity.

While the results of RST are very enlightening, it is still important to try to
understand the original CGHS equations. Large $N$ solubility was obtained
 in RST at the price of imposing an unnatural  global symmetry. One may
therefore wonder if the RST results are generic.
(Indeed, previous analyses have suggested that solutions of the CGHS equations
behave
 quite differently from those of the
RST equations.)
In fact we shall see in this
paper that most, although not all, of the features of the RST model appear in a
more generic setting.

A number of previous attempts have been made to analyse the CGHS equations,
but a definitive conclusion has been surprisingly elusive. It was originally
conjectured in CGHS that gravitational collapse did not lead to singularities.
This was subsequently shown to be false [\cite{BDDO,ORST}] in that black
hole-like singularities do arise, and static solutions corresponding to a
blackhole in equilibrium with a heat bath were found in [\cite{husk,nmr}].
It was also shown [\cite{ORST}] that if a large (apparent) black hole is
formed, its apparent horizon shrinks due to Hawking evaporation.
In [\cite{husk}] it was
suggested   that the apparent horizon and singularity
approach one another, meeting at future timelike infinity. In [\cite{thb}]
numerical evidence was given that they meet at a finite value of retarded time
at future null infinity. This cataclysmic phenomena was referred to as a
``thunderbolt".

In this paper we shall demonstrate numerically that in fact none of these final
outcomes are realized. Instead, we find that the apparent horizon and
singularity meet in a finite time,
in agreement with results of Lowe
[\cite{LOWE}]. A boundary condition is needed at the
junction point to  continue the evolution of the system. We argue that this
boundary condition can be chosen so that the system eventually settles back to
the vacuum. However, because  a spacelike singularity is present at
intermediate stages, information is lost in the process of large $N$ black
hole formation/evaporation.

Two-dimensional dilaton gravity coupled to $N$ matter fields $f_i$ is described
by the action
$$
S={1\over2\pi}\int
d^2\sigma\sqrt{-g}\left[e^{-2\phi}(R+4(\nabla\phi)^2+4\lambda^2)-{1\over2}
\sum^N_{i=1}(\nabla f_i)^2\right].\eqno(actn)
$$
We shall employ conformal gauge in which
$$
\eqalign{
g_{++}&=g_{--}=0,\cr
g_{+-}&=-{1\over2}e^{2\rho},\cr}
$$
where $\tau^\pm=\tau\pm\sigma$.
 In this gauge the ``linear dilaton" vacuum solution may be
written
$$
\eqalign{
\phi&={1\over2}(\tau^--\tau^+),\cr
\rho&=0, \cr}\eqno(ldv)
$$
where here and henceforth we choose units so that $\lambda=1$.
Classical black holes are formed by arbitrary infalling  ({\it i.e.\/},
$\partial_-f=0$)  matter from $\cI^-~~(\tau^-=-\infty)$.
The most general such classical solution is given by
$$
\eqalign{
f&=f(\tau^+),\cr
\phi&=-{1\over2}\ln\left[e^{\tau^+-\tau^-}-{1\over2}\int
 e^{\tau^+}\int e^{-\tau^+}\partial_+ f\partial_+ f \right],\cr
\rho&=\phi+{1\over2}(\tau^+-\tau^-),\cr}\eqno(csln)
$$
or a coordinate transformation thereof. The Penrose diagram
for this process is illustrated in
Figure~1.

At the quantum level black holes evaporate. In the large $N$ limit, with
$Ne^{2\phi}$ held fixed, the quantum theory is described by the semiclassical
CGHS equations. A convenient form of the $\rho-\phi$ equations is
$$
\eqalign{
8P\partial_+\partial_-\phi & = -P'
(4\partial_+\phi\partial_-\phi+\lambda^2e^{2\rho}),\cr
2P\partial_+\partial_-\rho & = e^{-4\phi}
(4\partial_+\phi\partial_-\phi+\lambda^2 e^{2\rho}),\cr} \eqno(rpeq)
$$
where
$$
\eqalign{
P & \equiv e^{-4\phi}-{N\over12} e^{-2\phi},\cr
P'& \equiv -4 e^{-4\phi}+{N\over6} e^{-2\phi}.\cr} \eqno(ppp)
$$
These two equations imply the constraint equations, which are given by
$$
 e^{-2\phi} (4\partial_+\phi\partial_+\rho-2\partial_+^2\phi)
+{1\over2} \sum^N_{i=1} \partial_+f_i\partial_+f_i
-{N\over12} (\partial_+\rho\partial_+\rho-\partial_+^2\rho)+
 t_+=0,\eqno(tpp)
$$
together with a similar equation for the $(--)$ components.
$t_+(t_-)$ is an arbitrary function of $\tau^+(\tau^-)$ determined by the
initial data.

 In interpretation of solutions
of (\call{rpeq}) it is useful to recall [\cite{CGHS,BDDO,dxbh}] that $1+1$
dimensional
dilaton gravity may be derived from $S$-wave reduction of $3+1$ dimensional
gravity. In that context $4\pi e^{-2\phi}$ measures the area of the two
spheres of constant radius [\cite{dxbh}].
Correspondingly a future (past) apparent horizon is a line along which
$\partial_+\phi (\partial_-\phi)$ vanishes[\cite{ORST}]. (This is distinct from
the event
horizon, which is a null line from behind which one cannot escape to $\cI^+$.)
This apparent horizon provides a useful two-dimensional definition of the
boundary of a black hole.

In the classical theory, the kinetic terms in the action degenerate at
$e^{-2\phi}=0$, which may be thought of as the origin since the two spheres
have zero area.
In the large $N$ quantum theory  this degeneration occurs at the
finite value $\phi_{cr}$ of $\phi$ such that $e^{-2\phi_{cr}}={N\over12}$, as
can be seen from inspection of (\call{rpeq}). Thus the quantum mechanical
origin is at
$\phi_{cr}$, and we shall not
try to evolve the system beyond this point.

We first compute the effects of sending in a shock wave of infalling
matter along $\tau^+=0$ with total energy $M$.
Initial value data for this problem may consist of the
values of $\rho$ and $\phi$ along two orthogonal null lines. We specify vacuum
data along $\tau^+=0$:
$$
\eqalign{
\rho(0,\tau^-)&=0,\cr
\phi(0,\tau^-)&={1\over2}\tau^-.\cr}\eqno(vdta)
$$
Along
$\cI^-$, the quantum equations exponentially approach the
classical equations, and the initial data should agree with the classical
solution
(\call{csln}) with ${1\over2}\partial_+f\partial_+f=M\delta(\tau^+)$:
$$
\eqalign{
\tau^+ & >0,\cr
\tau^- & \to -\infty,\cr
\phi \to & -{1\over2}\ln \left[M (1-e^{\tau^+}) +e^{\tau^+-\tau^-}\right],\cr
\rho \to & \phi +{1\over2}(\tau^+-\tau^-).\cr}\eqno(asdt)
$$
Of course it is not possible numerically to set initial data at
$\tau^-=-\infty$. Instead we specify the data at a large negative value
$\tau^-_0$ of $\tau^-$;
$$
\eqalign{
\tau^+ &>0,\cr
\phi_0 & =-{1\over2} \ln\left[M (1-e^{\tau^+}) +e^{\tau^+-\tau^-_0}\right],\cr
\rho_0 &= \phi+{1\over2} (\tau^+-\tau^-_0),\cr}\eqno(appr)
$$
which induces an error of order $e^{\tau^-_0}$. We then check numerically for
convergence at large  negative $\tau^-_0$.\footnote{\dag}{Erroneous
conclusions were reached in [\cite{thb}] as a result of choosing a small value
of
$\tau^-_0$.}

Our code approximates the spacetime using a null grid in $\tau^\pm$. It
computes the variables in some preset range $0<\tau^+<\tau^+_{MAX}$ and then
steps
up in $\tau^-$. The numerical computation cannot continue past a point where
$\phi=\phi_{cr}$, so it is instructed to revert to $\tau^+=0$ and integrate
along the next
$\tau^-$ step until the singularity is again
encountered. In this way it fills in the
region under (but not in the causal shadow of) the singularity.

The results are plotted
 for $M=.5$ and $N=12$ in figures 2a, 2b and 2c. Since $N$ may be scaled
out of the equations by shifting $\phi$ the
results are qualitatively independent
of $N$.  A black hole ({\it i.e.,} a region
where $\partial_+\phi>0$) forms a finite distance  along the infall line. A
spacelike singularity is produced when the shockwave reaches the origin where
$\phi=\phi_{cr}$ ($\phi_{cr}=0$ for $N=12$). All future-directed
ingoing null geodesics inside the black hole
terminate at the singularity. The apparent horizon becomes timelike
  due to Hawking
radiation.
Our computation clearly indicates that (contrary to previous claims) the
apparent horizon meets the singularity in a finite proper time. At this
``endpoint" denoted ($\tau^+_E,\tau^-_E$),
 the black hole has shrunk to zero size
(as measured by $e^{-2\phi}-{N\over12}$).

The endpoint  singularity is visible to observers outside the black hole with
coordinates
$\tau^\pm\geq\tau^\pm_E$, so this constitutes a violation of cosmic censorship.
Boundary conditions are
required  to evolve the system beyond the endpoint.
Physically one expects that it should be possible to impose boundary conditions
which restore the system to near-vacuum just after the endpoint. Indeed,
inspection of Figure~3 shows that along the edge of the causal shadow of the
singularity ($\tau^+>\tau^+_E,\tau^-_E$), the fields are very near their
 vacuum
values. However, they are not  {\it exactly\/} equal to their vacuum values.
This is also expected: the outgoing Hawking radiation itself induces particle
production.
 This leads to some flux crossing the line $(\tau^+>\tau^+_E,\tau^-_E)$
along which the black hole is seen to disappear.

Potentially this infalling energy flux could recollapse to form a new black
hole. However, we do not think that this is the case, because dilaton black
holes (of all sizes) radiate energy at a rate of order
$N/48$\footnote{\dag}{The
black hole mass $M$ is also of order $N$ with our conventions.}. Therefore
there is a
threshold for black hole production: energy must be thrown in faster than it is
radiated away. Such a threshold was found analytically in the RST model, and
numerical evidence for the existence of a similar threshold in the CGHS model
is illustrated in figures~4 and 5.

Although presumably no black hole is formed below the threshold, ({\it i.e.,\/}
$\partial_+\phi<0$ everywhere), the CGHS equations are still ill-defined at
$\phi=\phi_{cr}$. However since this is a timelike line, it is presumably
possible to choose reflecting boundary conditions so that the below-threshold
radiation passes through the origin without collapsing.

A plausible picture of black hole formation in the CGHS model is thus the
following. The black hole evaporates completely in a finite time and, with the
help of an endpoint boundary condition, disappears completely. Weak,
backscattered,
infalling radiation remains, but this can be reflected through the origin. The
excitations eventually dissipate and the system settles back down to the linear
dilation vacuum. The Penrose diagram is illustrated in Figure 6.

The CGHS model  differs from the RST model in this last respect: In the case of
RST, there is no
backscattered radiation, and the system is identically
 vacuum  in the
causal future of the black hole endpoint. This appears to be a non-generic
 feature
of the RST model, perhaps associated with the unnatural symmetries.

We have also considered the semiclassical one-loop ghost-corrected equations
derived in [\cite{fpbh}], and given by (\call{rpeq}) with the first equation in
(\call{ppp})
replaced by
$$
P=e^{-4\phi}-{N\over12} e^{-2\phi}+{N\over24}.\eqno(pfb)
$$
The extra term in (\call{pfb}) moves the critical values of $\phi$ into the
complex
plane and eliminates the singularity. The behavior of the system was analyzed
in [\cite{fpbh}]. In particular it was shown that the static solutions describe
black holes with non-singular, asymptotically deSitter interiors in thermal
equilibrium with a bath of Hawking radiation. A numerical simulation of shock
wave collapse described by these equations is illustrated in Figure~7. A more
detailed analysis is required to fully understand this system, but it appears
that the shock wave creates a black hole which subsequently evaporates down to
a stable remnant with a deSitter-like interior.
It is apparently possible to create a remnant with an arbitrarily weak shock
wave, which indicates that the remnant mass may be less than the linear dilaton
vacuum.

\centerline{Acknowledgements}
This work was supported in part by DOE grant DEAC-03-8ER4050, NSF grant
PHY89-04035 and the US-Israel BSF foundation. We are grateful to B. Birnir
and D. Goldwirth for useful conversations.
We also acknowledge the Aspen Center for Physics,
where this work was initiated.
\references

\refis{LOWE} D.A. Lowe, ``Semiclassical Approach to Black Hole Evaporation"
{\sl Phys. Rev.\/} {\bf D47},(1993) 2446.

\refis{CGHS} C.G. Callan, S.B. Giddings, J.A. Harvey and A. Strominger,
``Evanescent Black Holes", {\sl Phys.\ Rev.\/} {\bf D45\/}, R1005, (1992); For
recent reviews see J.A.~Harvey and A.~Strominger, ``Quantum Aspects of Black
Holes" preprint EFI-92-41, hepth@xxx/9209055, to appear in the proceedings of
the 1992 TASI Summer School in Boulder, Colorado, and S.B.~Giddings, ``Toy
Models for Black Holes Evaporation" preprint UCSBTH-92-36, hep-th@xxx/9209113,
to appear in the proceedings of the International Workshop of Theoretical
Physics, 6th Session, June 1992, Erice, Italy.

\refis{BDDO} T. Banks, A. Dabholkar, M.R. Douglas, and M. O'Loughlin, ``Are
horned particles the climax of Hawking evaporation?" {\sl Phys. Rev.\/} {\bf
D45}, (1992) 3607.

\refis{dxbh} S.B. Giddings and A. Strominger, ``Dynamics of Extremal Black
Holes", {\sl Phys. Rev.\/} {\bf D46\/}, (1992) 627.

\refis{fpbh} A. Strominger, ``Fadeev-Popov ghosts and $1+1$ dimensional black
hole evaporation", {\sl Phys. Rev. \/} {\bf D46\/}, (1992) 4396.

\refis{gs} S. Giddings and A. Strominger, ``Quantum theories of dilaton
gravity", {\sl Phys. Rev. \/} {\bf D46\/} (1993) 2454.

\refis{ORST} J.G. Russo, L. Susskind and L. Thorlacius,
``Black Hole Evaporation in $1+1$
Dimensions", \pl B292, 13, 1992.

\refis{thb} S.W. Hawking and J.M. Stewart, ``Naked and thunderbolt
singularities in black hole evaporation", hep-th/9207105.

\refis{AS} A. Strominger, unpublished.

\refis{RST} J.G. Russo, L. Susskind, and L. Thorlacius, {\sl Phys. Rev. \/}
{\bf D46\/} (1992) 3444.

\refis{husk} L. Susskind, and L. Thorlacius, ``Hawking radiation
and backreaction'', {\sl Nucl. Phys. \/} {\bf B382} (1992) 123.

\refis{nmr}S.W. Hawking, ``Evaporation of two dimensional black holes'', {\sl
Phys. Rev. Lett.\/} {\bf 69\/}, (1992) 406; B. Birnir, S. Giddings,
J. Harvey and A. Strominger, ``Quantum Black Holes'',  {\sl Phys. Rev. \/} {\bf
D46\/}, (1992) 638.

\refis{ABC} A. Bilal and C. Callan, ``Liouville models of black hole
evaporation", Princeton preprint PUPT-1320, hep-th/9205089; S.P. deAlwis,
``Quantization of a theory of 2d dilaton gravity", Boulder preprint
COLO-HEP-280, hep-th/9205069, ``Black hole physics from Liouville theory",
Boulder preprint COLO-HEP-284, hep-th/9206020, ``Quantum Black Holes in Two
Dimensions", Boulder preprint COLO-HEP-288, hep-th/9207095.

\endreferences
\endpage
\head{\bf Figure Captions}

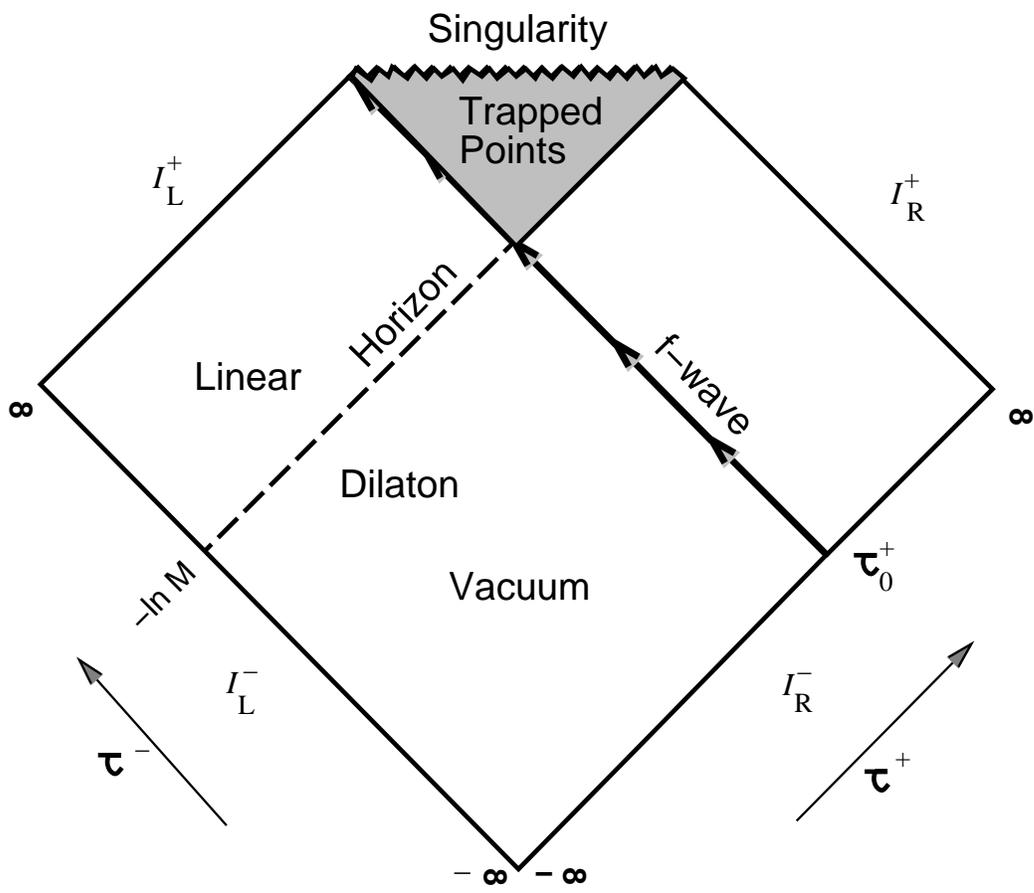
\noindent{\bf Figure 1.~~}Penrose diagram for classical black hole
formation by an $f$ shock wave with stress tensor
$\half\partial_+f\partial_+f=M\delta(\tau^+-\tau^+_0)$.

\noindent{\bf Figures 2a,~2b,~2c.~~}Numerical simulation of black hole
formation and evaporation. Initial conditions are specified
along the left and lower boundaries of the plot corresponding to
a null $M=.5$ shock wave along the left boundary.
The contours in Figure 2a depict lines of
constant $\phi$ (the rippled dashes are an artifact of the plotting
routine). The interior of the black hole is the region where
these lines slope downward to the
right, and the apparent horizon is the boundary of this region.
The results depicted here depend on two numerical
parameters: $\tau_0^-$, the initial $\tau^-$ value, and $d\tau$, the
numerical time step. In order to ensure that the results are not due
to numerical errors we checked the convergence of our results as
$\tau_0^- \rightarrow -\infty$ and $d\tau \rightarrow 0$. In Figure 2b we
depict the singularity line $\phi=\phi_{cr}$ and the apparent
horizon line $\partial_+ \phi = 0$ for  $d\tau$
values  ranging between $4 \cdot 10^{-3}$ and $6.25 \cdot 10^{-5}$.
It is evident that the curves converge. More quantitative
tests show that the errors decrease like $d\tau^2$ and $e^{\tau^-_0}$,
as expected.

Figure 2b suggests that singularity and the horizon intersect.  However,
technically, our scheme cannot reach the intersection point
$(\tau^+_E,\tau^-_E)$ itself.  The numerical scheme breaks down and it
is impossible to continue the computation in the $\tau^+$ direction
when $\phi=\phi_{cr}$.  Regardless of how fine our resolution is we
can never reach the endpoint
$(\tau^+_E,\tau^-_E)$.  There will always be a gap
of one computation step $d\tau$ between the last $\tau^-$ value for
which we calculated the horizon's position and the first $\tau^-$ that
is singular.  However, Figure 2c shows an enlargement of the region near
$(\tau^+_E,\tau^-_E)$. The different points correspond to the two
finest (smallest $d\tau$) computations and the curves describe a Pade
extrapolation through these points. The agreement between the computed
points demonstrates the numerical convergence of our computation. The
fact that the extrapolated curves intersect, at a point just beyond
the end of the computation, is clear evidence that
$(\tau^+_E,\tau^-_E)$ exist and are near $(.3154,-.1093)$.

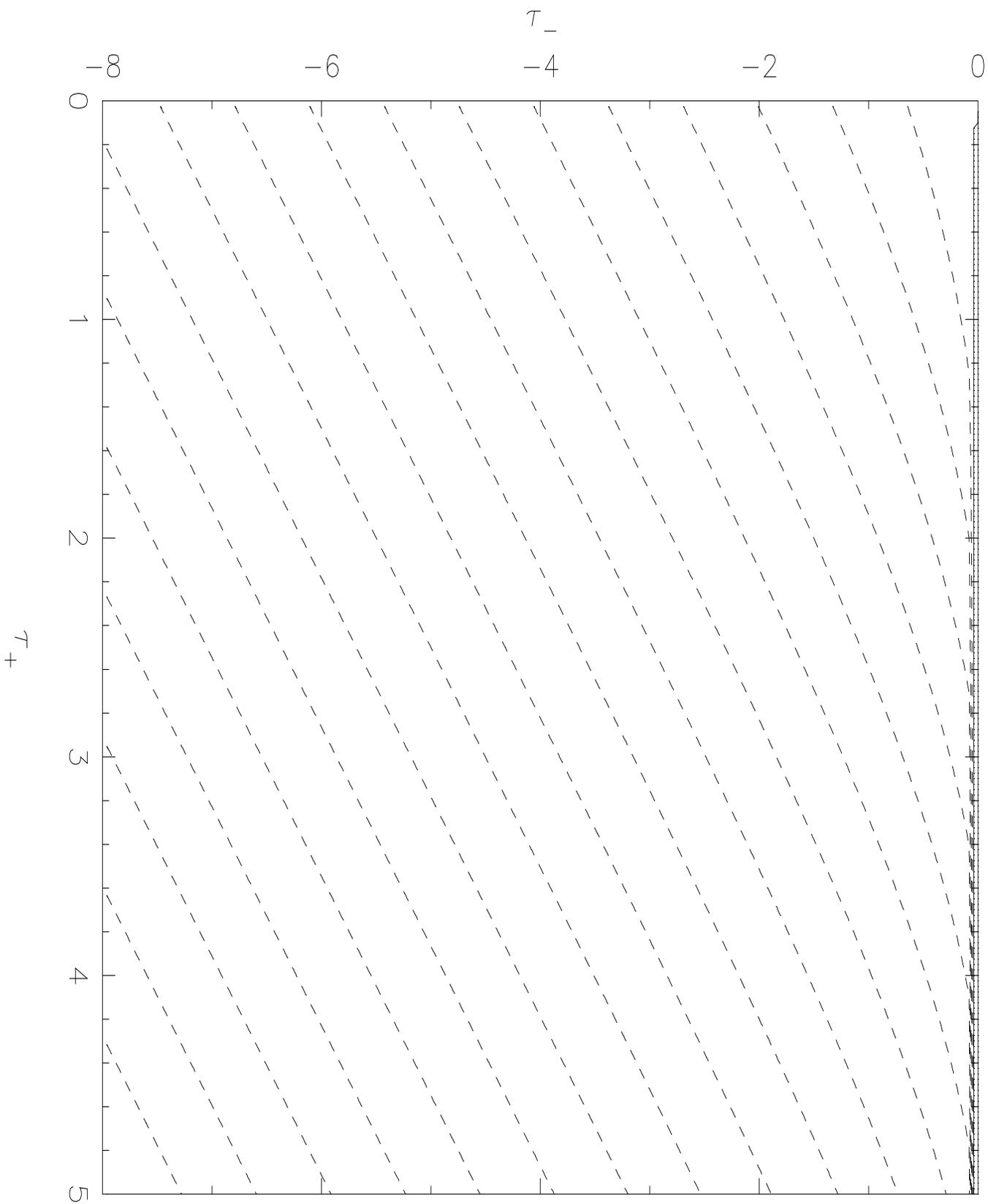
\noindent{\bf Figure 3.~~}Numerical simulation of black hole
formation and evaporation. Initial conditions are specified
along the left and lower boundaries of the plot corresponding to
a null $M=.5$ shock wave along the left boundary. This plot
extends to larger values of $
\tau^+$ than those appearing in
Figure 2a. It is evident that lines of constant
$\phi$ cross into the causal shadow of the black hole endpoint
along $\tau^-_E\sim -0.1$. The nearly straight
nature of those lines for large $\tau^+$ also indicates that
the system is near its vacuum configuration far from the black hole.

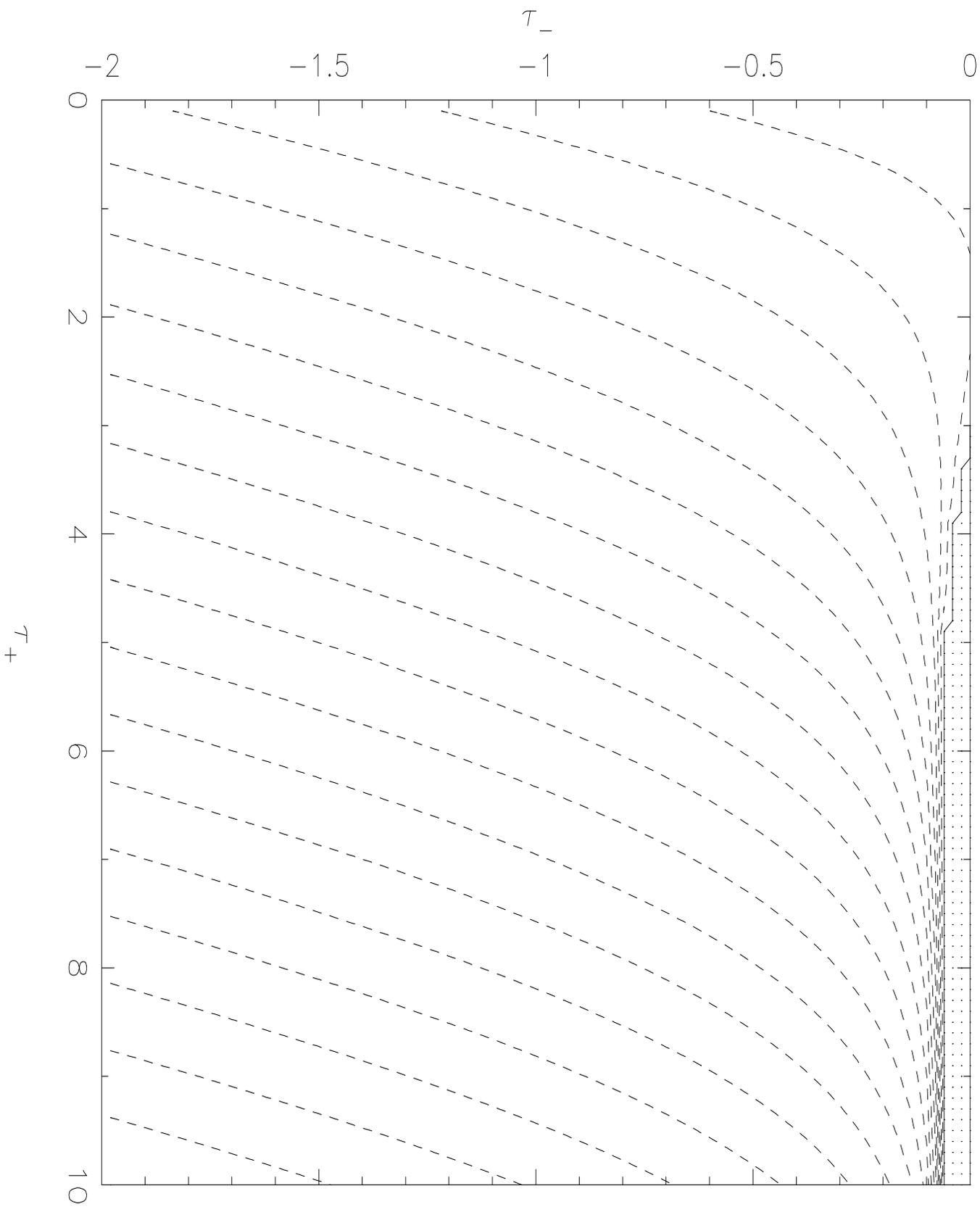
\noindent{\bf Figure 4.~~}Energy is thrown towards the origin at a constant,
above-threshhold rate
for $\tau^+>0$, corresponding to a
matter stress tensor $\half \partial_+f  \partial_+f=\theta(\tau^+)$.
A black hole forms and grows because matter accretion outpaces
Hawking radiation. A spacelike singularity eventually crosses the
null line $\tau^-=0$ where the numerical simulation ends.

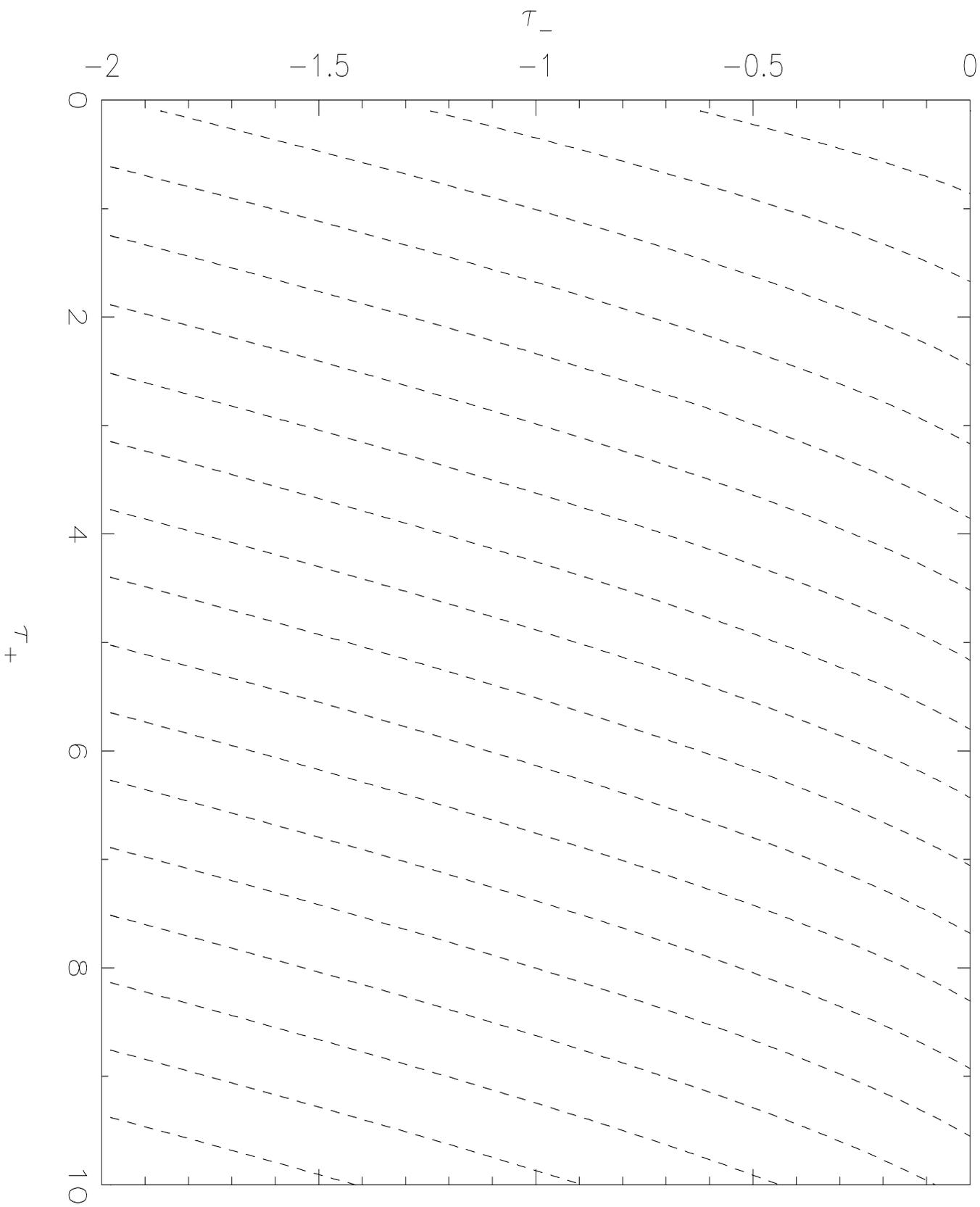
\noindent{\bf Figure 5.~~}Energy is thrown towards the origin at a constant,
below-threshhold rate
for $\tau^+>0$,corresponding to a
matter stress tensor $\half\partial_+f\partial_+f=.5\theta(\tau^+)$.
A black hole does not form because Hawking radiation outpaces
matter accretion. No singularity crosses the
null line $\tau^-=0$ where the numerical simulation ends.

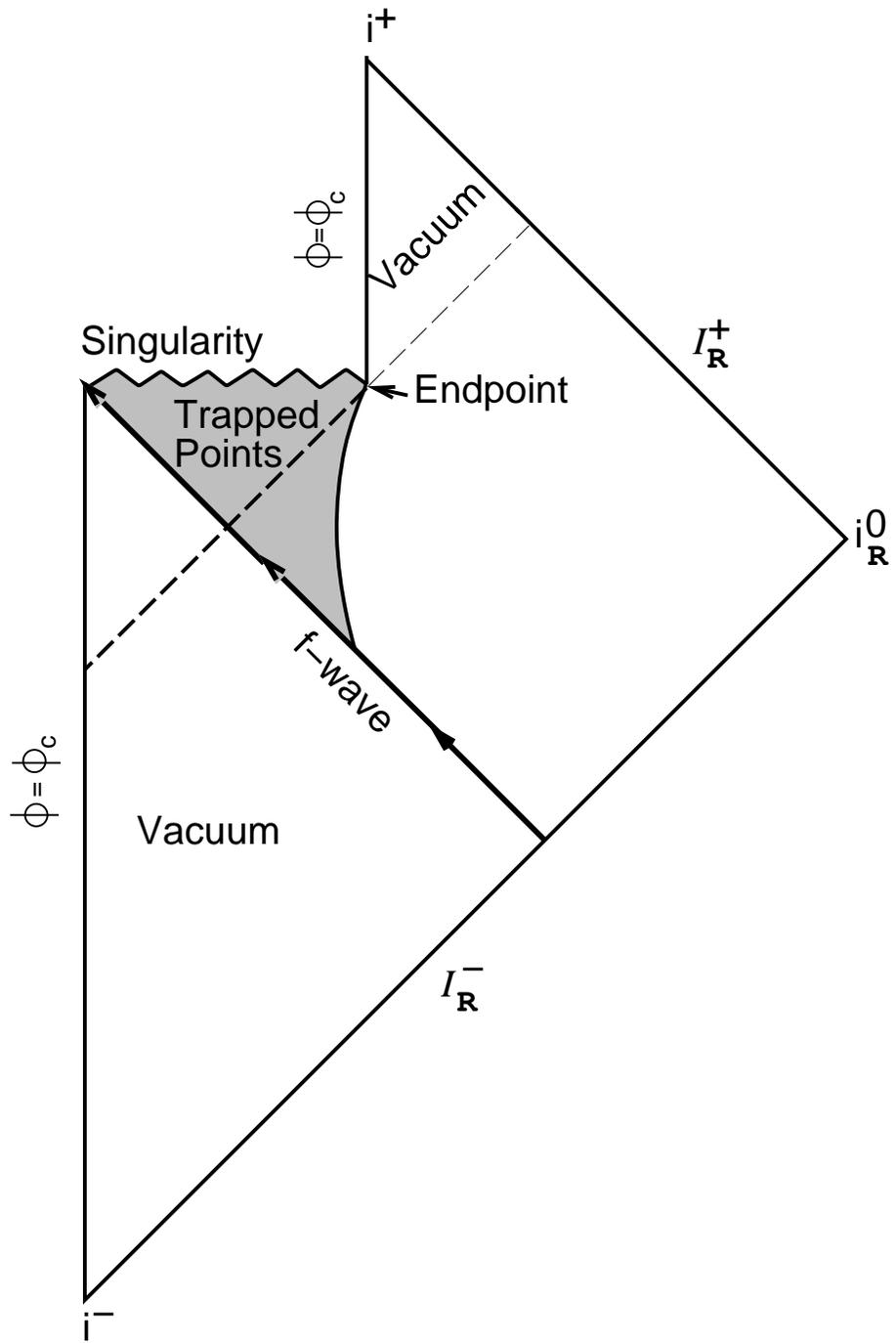
\noindent{\bf Figure 6.~~}Conjectured Penrose diagram for
large-$N$ black hole formation/evaporation. The black hole disappears
in a finite time. Boundary conditions can be chosen along the
timelike segment of the line $\phi = \phi_{cr}$ after the endpoint in
such a way that the system eventually settles back down to the vacuum.

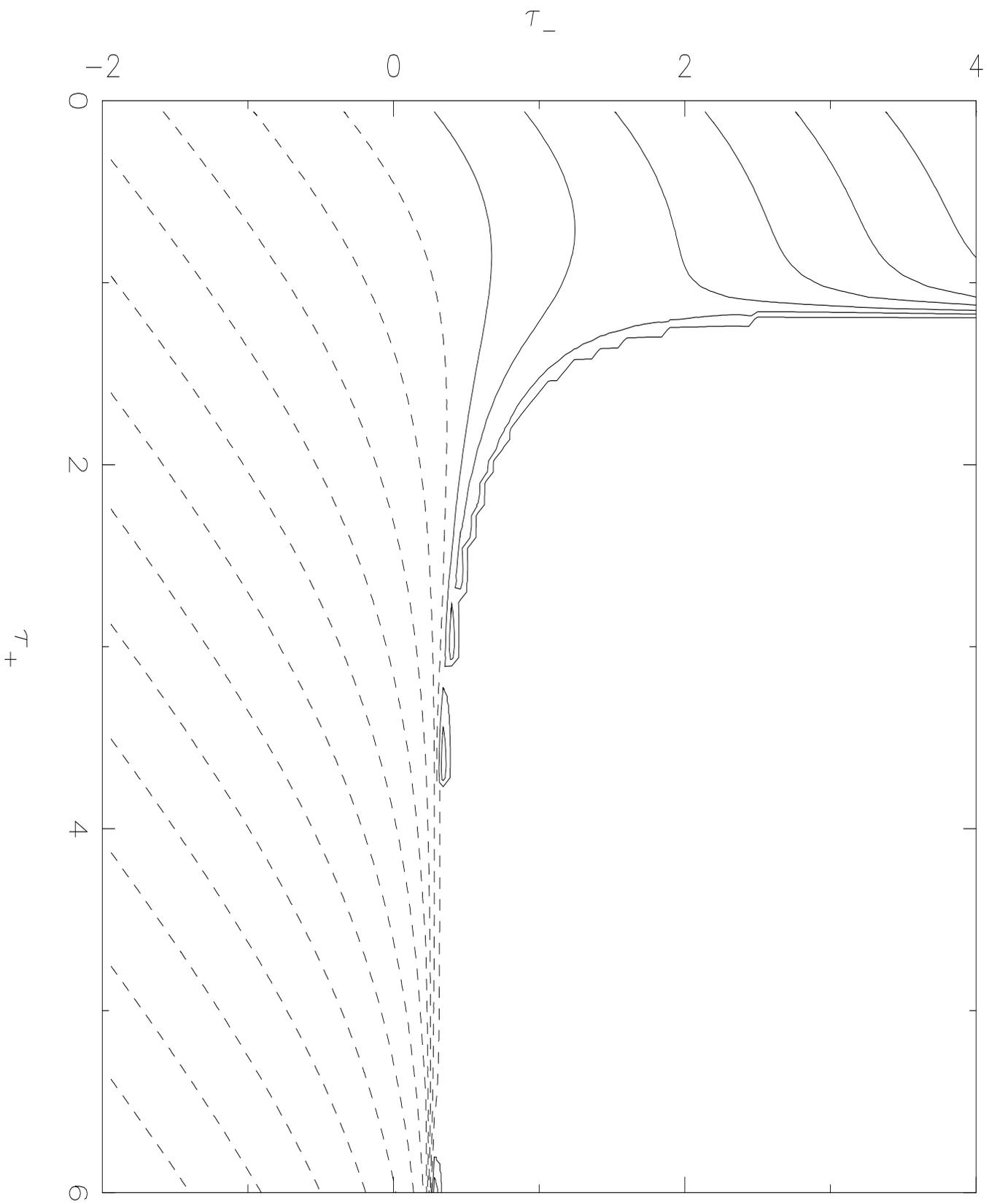
\noindent{\bf Figure 7.~~}Numerical simulation of black hole
formation and evaporation, as described by the one-loop ghost-corrected
semiclassical equations with $N=12$. Initial conditions are specified
along the left and lower boundaries of the plot corresponding to
a null $M=.1$ shock wave along the left boundary.
The contours depict lines of
constant $\phi$. The simulation is terminated when $e^{2\rho}$
diverges. This is apparently a coordinate divergence an
infinite distance away from finite points in the
spacetime, indicating the possible formation of
a deSitter-like region inside the horizon. For large
$\tau^+$ the system appears to settle down to a static remnant.
\endit
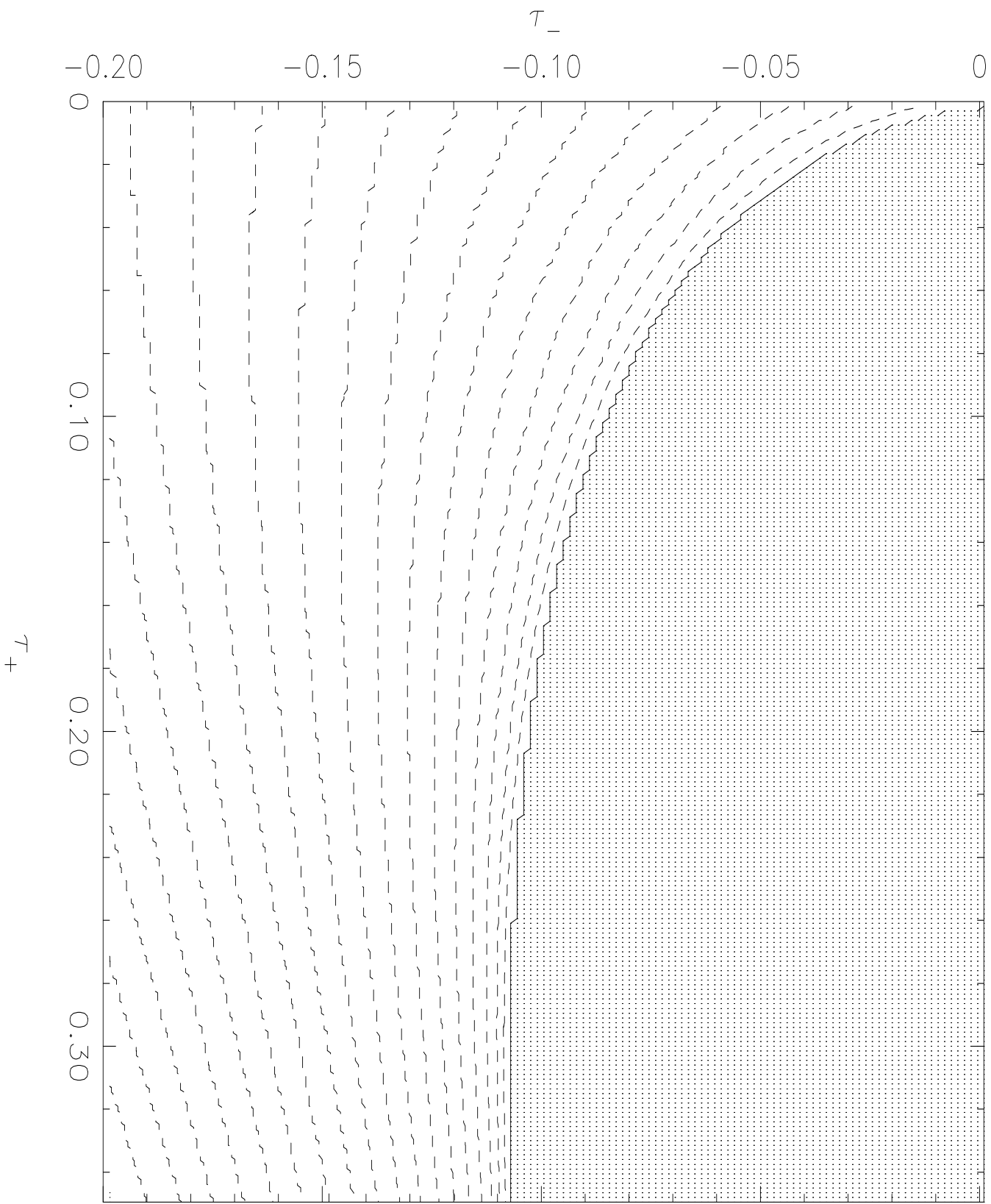
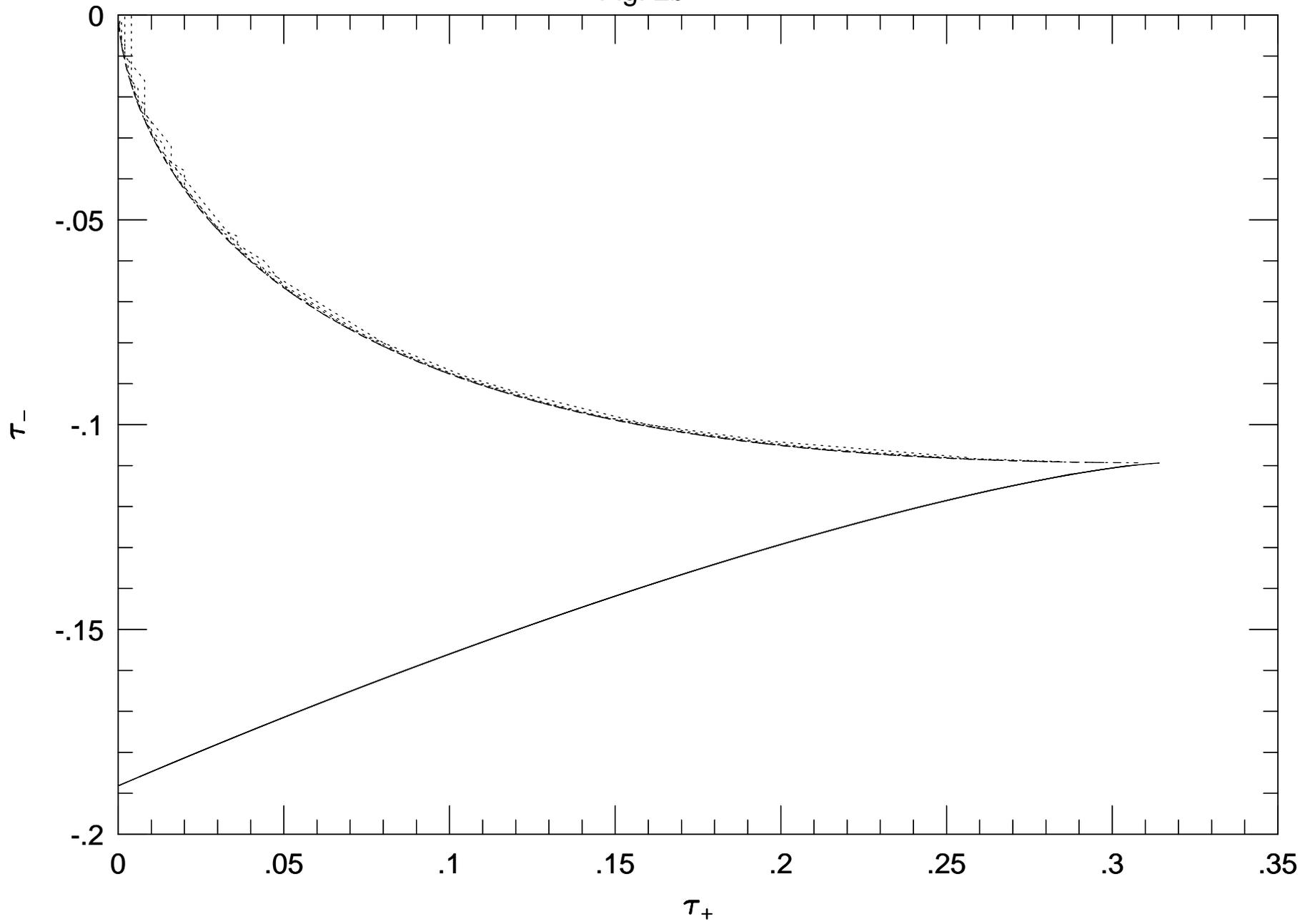
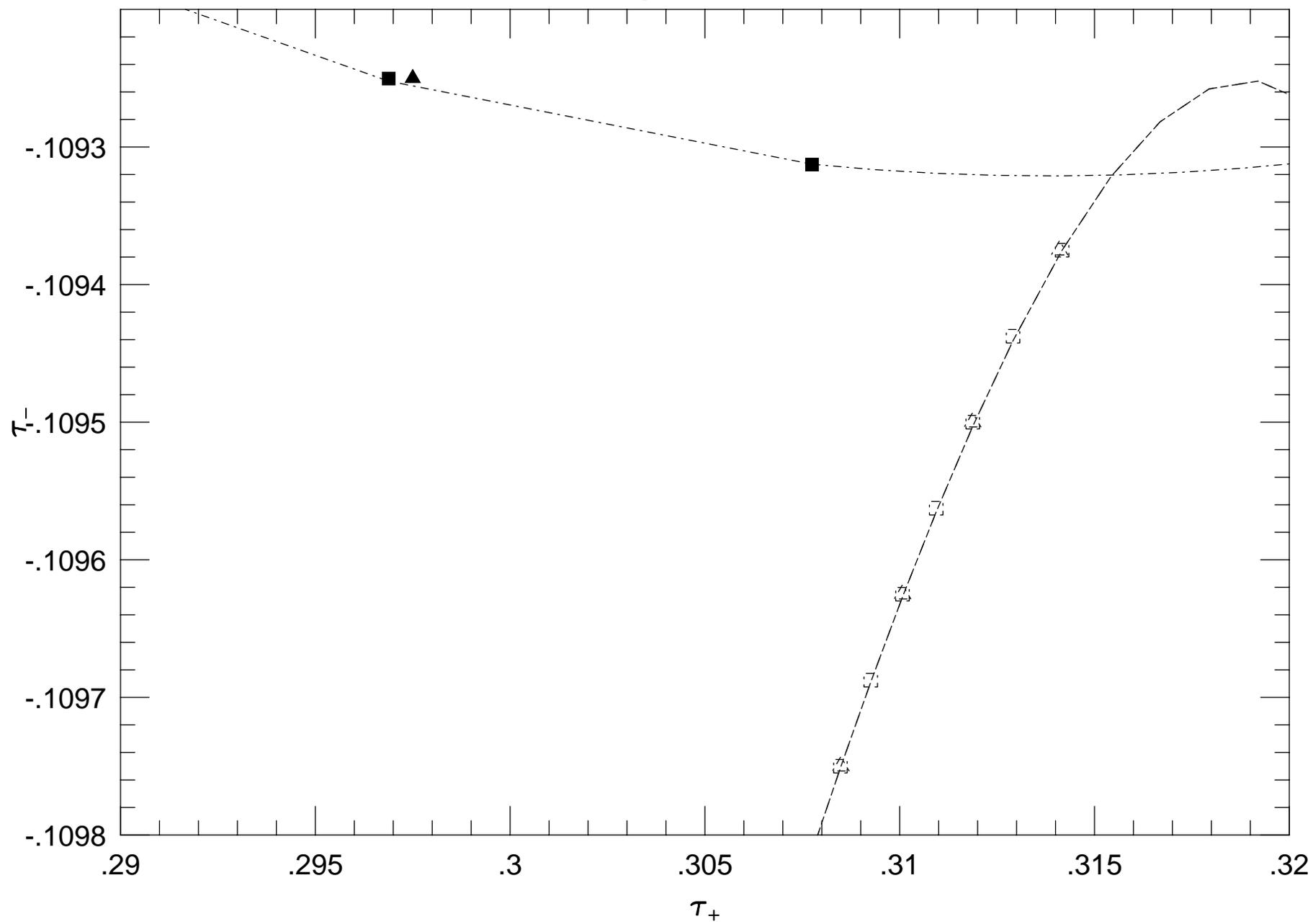
